\documentclass[aps,pra,twocolumn,preprintnumbers,showpacs,superscriptaddress,amsmath,amssymb]{revtex4}
\usepackage{dcolumn}
\usepackage{bm}
\usepackage{amssymb}
\usepackage[german, english]{babel}
\usepackage{graphicx}
\usepackage{color}
\usepackage{textcomp}
\usepackage[letterpaper,total={7in,9.5in},top=0.75in,left=0.75in]{geometry}

\usepackage{array}

\begin{document}
\title{On an attempt to optically excite the nuclear isomer in Th-229}

\author{Simon Stellmer}
\affiliation{Atominstitut, TU Wien, 1020 Vienna, Austria}
\author{Georgy Kazakov}
\affiliation{Atominstitut, TU Wien, 1020 Vienna, Austria}
\affiliation{Wolfgang Pauli Institute, 1090 Vienna, Austria}
\author{Matthias Schreitl}
\affiliation{Atominstitut, TU Wien, 1020 Vienna, Austria}
\author{Hendrik Kaser}
\affiliation{Physikalisch-Technische Bundesanstalt, 10587 Berlin, Germany}
\author{Michael Kolbe}
\affiliation{Physikalisch-Technische Bundesanstalt, 10587 Berlin, Germany}
\author{Thorsten Schumm}
\affiliation{Atominstitut, TU Wien, 1020 Vienna, Austria}

\date{\today}

\pacs{06.30.Ft, 78.20.-e}

\begin{abstract}
We aim to perform direct optical spectroscopy of the $^{229}$Th nuclear isomer to measure its energy and lifetime, and to demonstrate optical coupling to the nucleus. To this end, we develop $^{229}$Th-doped CaF$_2$ crystals, which are transparent at the anticipated isomer wavelength. Such crystals are illuminated by tunable VUV undulator radiation for direct excitation of the isomer. We scan a $\pm 5\,\sigma$ region around the assumed isomer energy of 7.8(5)\,eV and vary the excitation time in sequential scans between 30 and 600 seconds. Suffering from an unforeseen strong photoluminescence of the crystal, the experiment is sensitive only to radiative isomer lifetimes between 0.2 and 1.1 seconds. For this parameter range, and assuming radiative decay as the dominant de-excitation channel, we can exclude an isomer with energy between 7.5 and 10 eV at the 95\% confidence level.
\end{abstract}

\maketitle

\section{Introduction}
\label{sec:introduction}

The typical energy scales in nuclear physics on one side and in atomic physics on the other side are vastly different. While the energy released in radioactive decay is usually a few MeV and the spacing between excited nuclear states is on the order of 100\,keV, the excitation and ionization energies of valence electrons in neutral atoms are only a few eV. Similarly, the typical bandgap in semiconductors is also a few eV, and the wavelength region covered by established laser technology corresponds to a few eV as well.

Among the 176,000 nuclear states currently charted for all known isotopes, only a single one is known which has an excitation energy in the range of electronic excitations and current laser technology:~the famous isomeric state in $^{229}$Th \cite{Kroger1976fot,Reich1990eso,Helmer1994aes,Beck2007eso,Beck2009ivf,Matinyan1998laa,Dykhne1998meo,Peik2003nls,Flambaum2006eeo,Berengut2009pem,Rellergert2010cte,Campbell2011wco,Porsev2010eot,Campbell2012sin,Kazakov2012poa,Kazakov2014pfm,Herrera2014elo,Jeet2015roa,Tkalya2015rla,Yamaguchi2015esf,Wense2015ddo,Seiferle2017fso,Seiferle2017lmo,Wense2017ale,Minkov2017rtp,Thielking2018lsc}; see \textit{e.g.}~Refs.~\cite{Peik2015ncb,Wense2018otd} for recent reviews. The current best determination of its energy is inferred from indirect gamma spectroscopy \cite{Kroger1976fot,Reich1990eso,Helmer1994aes,Beck2007eso} and suggests a value of 7.8(5)\,eV \cite{Beck2007eso,Beck2009ivf}, corresponding to a wavelength of 160(10)\,nm. An isomer energy range of 6.3\,eV to 18.3\,eV has been confirmed by a recent experiment, in which the existence of the isomer was proven through the detection of internal conversion (IC) electrons \cite{Wense2015ddo}. While there is already a measurement of the isomeric half-life under de-excitation via IC \cite{Seiferle2017lmo}, there exist only theoretical calculations of the radiative half-life \cite{Helmer1994aes,Dykhne1998meo,Ruchowska2006nso,Tkalya2015rla,Minkov2017rtp}. These estimates span a range between 1000\,s and 40\,000\,s.

Owing to the low energy and long lifetime of its isomer, $^{229}$Th has been suggested as a candidate for a \emph{nuclear} optical clock \cite{Peik2003nls,Campbell2012sin,Kazakov2012poa}, for a measurement of temporal variations in fundamental constants \cite{Flambaum2006eeo,Berengut2009pem,Rellergert2010cte}, and for nuclear quantum optics \cite{Burvenich2006nqo}. All of these applications rely on an optical excitation of the isomer $^{229\mathrm{m}}$Th from the nuclear ground state. Aside from electron bridge processes, which rely on the coupling of the electronic shell to the nucleus to enhance the excitation probability \cite{Porsev2010eot}, direct optical excitation of the isomer is the ``holy grail'' of contemporary $^{229}$Th research \cite{Jeet2015roa,Yamaguchi2015esf,Wense2017ale}.

Optical spectroscopy is required to improve on the uncertainty of the isomer energy, which can only be determined to within about 0.1\,eV using available nuclear physics methods, such as advanced gamma spectroscopy with magnetic micro-calorimeters \cite{Kazakov2014pfm} or spectroscopy of the IC electron \cite{Seiferle2017fso,Wense2017ale}. Note that the first ionization threshold of the neutral Th atom is below the isomer energy, leading to fast IC decay of neutral $^{229\mathrm{m}}$Th. This decay channel is heavily suppressed in higher charge states \cite{Wense2015ddo}.

Experiments employing optical excitation of nuclei can follow two strategies: (1) ion trapping of some $10^6$ $^{229}$Th ions \cite{Campbell2011wco,Herrera2014elo} combined with direct VUV excitation or electron bridge processes, (2) deployment of a thin metallic sample, in which the isomer undergoes rapid IC, and the IC electrons are detected \cite{Wense2017ale}, or (3) usage of a large-bandgap material, into (or onto) which $^{229}$Th ions are doped (adsorbed) such that the IC channel is suppressed by the absence of low-energy electronic states in the microscopic environment of the Th ion, and detection of the gamma ray emitted upon de-excitation \cite{Jeet2015roa,Yamaguchi2015esf}. In the study presented here, we follow the latter approach.

Two experiments attempting to optically excite the isomer have already been reported. In the first experiment, $^{229}$Th ions doped into LiSrAlF${}_6$ crystals were illuminated at the Advanced Light Source (ALS) in Berkeley, USA. An energy range of about half the estimated uncertainty range around 7.8(5)\,eV was covered, but no signal was found \cite{Jeet2015roa,Tkalya2015rla}. In the other experiment, $^{229}$Th was adsorbed onto a thin CaF$_2$ crystal substrate and illuminated at the Metrology Light Source (MLS) in Berlin, Germany, with a scan range of 3.9\,eV to 9.5\,eV, without detection of a resonance signal \cite{Yamaguchi2015esf}. It is not known whether the isomer energy and lifetime were outside the parameter ranges scanned in these two experiments, or whether the IC decay channel also dominates in these samples, thereby quenching the radiative channel.

In this paper, we present a new attempt in this direction. We employ Th-doped CaF$_2$ crystals, which have a lattice structure accessible to \textit{ab initio} density functional calculations \cite{Dessovic2014tdc}. The doping efficiency nears unity, and Th${}^{4+}$ ions distribute almost homogeneously in the crystal. Earlier work has already characterized this system in terms of photo- and radioluminescence \cite{Stellmer2015rap,Stellmer2016fso}. In the present study, we use VUV radiation of the MLS undulator to cover a spectral range of $\pm5\sigma$ around the anticipated isomer energy.

The remainder of this paper is structured as follows: in Sec.~\ref{sec:setup}, we use parameters of the Th-doped crystal, the excitation light source, and the detection system to estimate the expected signal and background count rates. The measurement is presented in Sec.~\ref{sec:measurements}. In the following Sec.~\ref{sec:analysis}, we detail an elaborate analysis to construct a parameter space of isomer energy and lifetime that can be excluded on the basis of our measurement. A conclusion is given in Sec.~\ref{sec:conclusion}.

\section{Experimental setup}
\label{sec:setup}

\subsection{Th-doped crystal}

We choose CaF$_2$ as the host crystal for two reasons: (1) it is a well-established material in VUV optics, and (2) the similarity of the ionic radii of Ca$^{2+}$ and Th$^{4+}$, leading to a high doping efficiency. Atomistic model calculations show that Th$^{4+}$ ions replace Ca$^{2+}$ ions in the crystal lattice, with two associated fluorene interstitials (F$^{-}$) for charge compensation \cite{Dessovic2014tdc}. The bandgap in CaF$_2$ is about 10\,eV \cite{Heaton1980eeb}.

Th-doped crystals are grown using the vertical gradient freeze technique with a scavenger process applied to remove oxygen from the system. A seed crystal is used to set the crystal orientation, and we grow single crystals in the 1-1-1 orientation. Raw crystals have a cylindrical shape (17\,mm in diameter, 40\,mm height) and are cut and polished to the desired geometric shape.

The $^{229}$Th material, typically a few $10\,\mu$g per crystal, is blended with the 100-fold quantity of the stable isotope $^{232}$Th to generate a quantity that can easily be handled. Before the growth process, the Th material is added to the CaF$_2$ powder in the form of ThF$_4$, and a CaF$_2$ seed crystal is used to set the crystal axes. The doping efficiency of Th into CaF$_2$ is about unity. We also characterize the macroscopic distribution of Th in the crystal with a resolution of about 3\,mm. This is done by cutting a crystal into a set of mm-sized pieces and performing neutron activation analysis on them. Both the vertical and radial distribution of Th are nearly uniform, with at most a factor of 1.5 difference between the top/bottom and inner/outer regions of the raw crystal. The microscopic environment of the Th dopant has not yet been investigated.

The specific crystal used in this study has a length of 13.15\,mm and a doping concentration of $n_{\mathrm{Th}} = 8.8(5) \times 10^{15}\,\mathrm{cm}^{-3}$, corresponding to a number of $3.3(2)\times 10^{14}$ nuclei. The $^{229}$Th alpha activity is 900(50)\,Bq to comply with radiation safety regulations, as 1\,kBq is the exemption limit in Austria and Germany. Only a fraction of $3.7(2) \times 10^{-7}$ of all Ca${}^{2+}$ ions are replaced by $^{229}$Th${}^{4+}$ ions.

The key parameter for nuclear forward scattering (NFS) \cite{Burck1992nfs,Peik2003nls} is the amount of nuclei within a volume of $\lambda^3$, where $\lambda$ is the excitation wavelength. For $\lambda = 160\,\mathrm{nm}$, the density of nuclei is $n_{\mathrm{Th}} = 36(2)/\lambda^3$. The critical density of nuclei is strongly reduced by inhomogeneous and homogeneous (decoherence) broadening effects, and it is justified to neglect collective effects for the study presented here. 

The facets of the crystal have a wedge of less than 0.1\,mrad. For the wavelength-dependent refractive index of $\rm CaF_2$, we use the relation
\begin{equation}
n(\lambda)=1.405 + \frac{9.69}{\lambda - 94},
\end{equation}
where $\lambda$ is given in nanometers. The crystals are transparent down to about 125\,nm.

\subsection{Light source}

A beamline station has been installed at the U125 undulator beamline of the MLS synchrotron in Berlin, Germany \cite{Gottwald2012cca}. A stored electron beam (maximal nominal current 200\,mA, energy 629\,MeV, typical lifetime 6 hours) is passed through an undulator structure with $N=30.5$ periods to gain undulator radiation. The output wavelength of the first harmonic can be tuned between 65\,nm and 1\,$\mu$m.

The nominal FWHM linewidth $\Delta E$ at photon energy $E$ is $\Delta E = 2E/N$, corresponding to about 0.53\,eV at 8\,eV photon energy (10.6\,nm at 160\,nm nominal wavelength). This linewidth has been verified experimentally; see Fig.~\ref{fig:fig2}(b). For the analysis described below, the excitation spectrum is parametrized by
\begin{equation}
\begin{split}
  P_s(\lambda,\lambda_0) &=\frac{1}{N_{\gamma}} \frac{dN_{\gamma}}{d\lambda} \\
     &= A\frac{\exp\left[-\frac{\lambda-\lambda_m}{2 \Gamma^2}\right]
             }
             {1+\left(\frac{\lambda-\lambda_m}{\Sigma}\right)^2
             }
             (1+\tanh[\alpha(\lambda-\lambda_m)]).
\end{split}
\label{eq:excitation_spectrum}
\end{equation}
The fit parameters $A$, $\lambda_m$, $\Gamma$, $\Sigma$, and $\alpha$ are determined for five values of the nominal wavelength $\lambda_0$. Afterwards, an interpolation is performed for all other values of $\lambda_0$ used in the experiment.

The undulator radiation also contains higher harmonics at $n$-times the nominal photon energy, which lead to damage of the transmissive optics in the beam path. To filter these higher harmonics, the beam passes a periscope of two gold-coated copper mirrors, the first of them is heated to 100${}^{\circ}$C to avoid contaminations. This periscope is also used for vertical deflection of the beam.

The beam has a focus at the position of the crystal (horizontal beam diameter $s$ about 1.0\,mm, vertical 0.5\,mm). The light intensity at the position of the crystal is measured and constantly monitored by a calibrated Cs-I photocathode, placed about 10\,cm downstream of the crystal; see Fig.~\ref{fig:fig1}. For a beam current of 100\,mA, the total photon flux has been measured to be $4.7(2) \times 10^{14}\,\mathrm{s}^{-1}$ between 130 and 170\,nm, which corresponds to the plateau of maximum quantum efficiency of the Cs-I photocathode; see Fig.~\ref{fig:fig2}(a). In our analysis below, we will conservatively assume a flux of $4.0 \times 10^{14}\,\mathrm{s}^{-1}$ over the entire wavelength range studied here. This corresponds to a peak spectral flux of about $3\,$ photons/(Hz$\times$s) and a total intensity of 0.5\,mW.

\subsection{Detection system}

\begin{figure}
  \centering
  \includegraphics[width=\columnwidth]{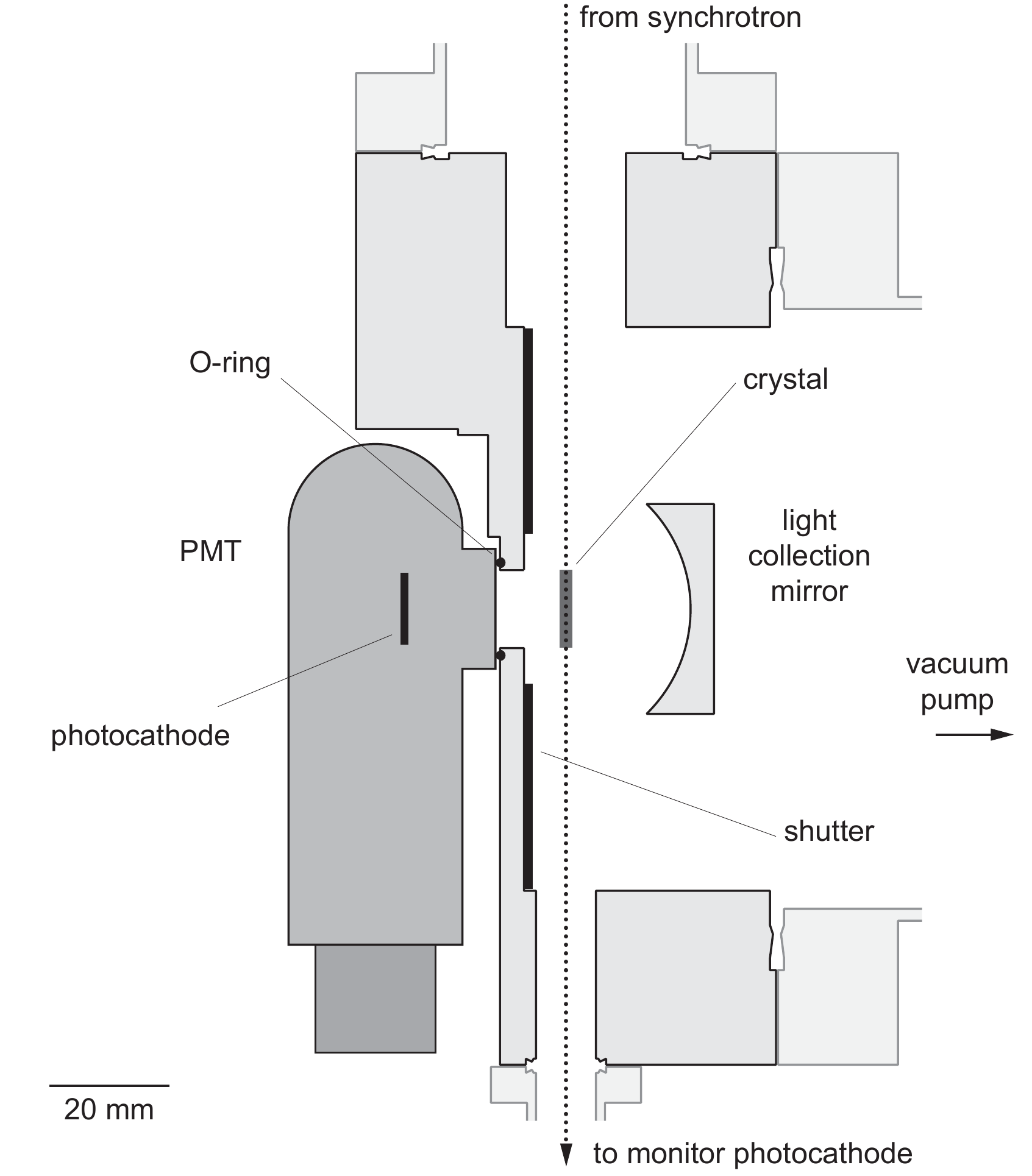}
  \caption{The experimental setup, shown here as a cut in the horizontal plane which includes the path of the excitation beam (dotted line).}
  \label{fig:fig1}
\end{figure}

The detection system is arranged around the crystal; see Fig.~\ref{fig:fig1}. A spherical mirror with Al+MgF$_2$ coating (diameter $D=35\,\mathrm{mm}$, curvature $r_{cc}=30\,\mathrm{mm}$, VUV reflectivity $R\sim80\%$) is used to image the crystal emission onto the photocathode of a side-on photo-multiplier tube (PMT). The PMT is protected from stray light during crystal illumination by a mechanical shutter. The MgF$_2$ window of the PMT is directly sealed to the vacuum by a FFKM O-ring \cite{Yamaguchi2015esf}. The distance between the crystal and the photocathode is about 27\,mm, leading to a de-magnification of about 0.5. The size of the photocathode is $8 \times 12\,\mathrm{mm}^2$ ($3 \times 12\,\mathrm{mm}^2$ for the diamond photocathode, see below), such that the total geometric collection efficiency is about 8\%.

Three different types of PMTs are used for measurements, namely Hamamatsu models R7639 (diamond, 115 to 230\,nm), as well as R8487 and R10454 (Cs-I, 115 to 195\,nm); see Fig.~\ref{fig:fig2}(c) for wavelength-dependent quantum efficiencies. In addition, PMTs model R8486 (Cs-Te photocathode, sensitivity range 115 to 320\,nm) are used for characterization measurements. Note that all of these photocathodes have a residual sensitivity at longer wavelengths, on the order of $10^{-3}$ times the maximum quantum efficiency. Given the massive crystal photoluminescence at around 300\,nm \cite{Stellmer2016fso}, this residual sensitivity cannot be neglected in our measurements. For the Cs-I model, we employ so-called extra solar-blind PMTs with reduced contaminations of alkali atoms on the photocathode, leading to a five-fold reduction of sensitivity to light in the visible range (R10454). Typical dark count rates range between 3 and 25 counts per second (csp) for the different models, which are negligible compared to the crystal luminescence background.

The PMT signal is amplified and sent to a Becker \& Hickl PMS-400-A multiscaler for counting of the pulses. The multiscaler card is synchronized to the closing of the beam shutter, with a delay of 200\,ms added to ensure that no excitation light blinds the PMT. The counts are saved in bins of typically 100\,ms length, along with monitored values of the beam current and the excitation light power. Later, all data sets will be normalized to the average beam current during the preceding excitation time. 

\subsection{Expected signal rate}

The experiment described here is sensitive only to radiative de-excitation of the isomer, and consequently, we suppose that the only relaxation channel of the $\rm ^{229\mathrm{m}}Th$ isomer in the $\rm CaF_2$ crystal lattice environment is the radiative decay. Also, we suppose that the isomer transition is a pure M1 magnetic dipole transition, and, as it was shown in \cite{Tkalya2000sep}, its radiative decay rate is enhanced by a factor of $n^3$ in comparison with the decay rate $\Gamma_{sp}$ of the bare nucleus. Here, $n=n(\lambda_{is})$ is a refractive index of the crystal corresponding to the (vacuum) wavelength $\lambda_{is}$ of the isomer transition. The number $N_{is}$ of Th nuclei in the isomeric state after excitation time $t_{exc}$ may be expressed as \cite{Kazakov2012poa,Jeet2015roa,Yamaguchi2015esf,Wense2017ale}
\begin{equation}
N_{is}= N_{eq} \left(1-e^{-\Gamma_{sp} n^3 t_{exc}}\right),
\label{eq:3}
\end{equation}
where
\begin{equation}
N_{eq} = \frac{N_{Th} W}{\Gamma_{sp} n^3} =
\frac{N_{Th}}{\Gamma_{sp} n^3} \frac{\lambda_{is}^5 \Gamma_{sp}}{24 \pi^2 c^2 \hbar} \frac{dP_{\gamma}}{s\, d\lambda}=
\frac{n_{Th} \ell \lambda_{is}^4}{12 \pi \, c \, n^3} \frac{dN_\gamma}{d\lambda \, dt}.
\end{equation}
$N_{\mathrm{Th}}$ is the number of Th nuclei irradiated by the excitation light beam, ${dP_\gamma}/{d\lambda}$ is the power of this beam in the crystal per unit wavelength, $N_{eq}$ is an ``equilibrium'' number of excited nuclei after long-term irradiation, $W$ is the excitation rate, $\lambda_{is}$ is the wavelength of the isomer transition, $dN_{\gamma}/(dt d\lambda)$ is the photon flux per unit wavelength per second, $c$ is the speed of light, $\hbar$ is the Planck constant divided by $2\pi$, and $\ell$ is the length of the crystal. Expressing $dN_{\gamma}/(dt d\lambda)$ via the nominal wavelength $\lambda_0$, isomer wavelength $\lambda_{is}$, and storage ring current $I_0$, with the help of equation (\ref{eq:excitation_spectrum}) we obtain
\begin{equation}
N_{eq}=\frac{n_{Th} \ell \lambda_{is}^4}{12 \pi \, c \, n^3} P_s(\lambda_{is}, \lambda_0)\,  P_I (\lambda_0) \cdot I_0.
\label{eq:4}
\end{equation}

The average number $y_k^{is}$ of counts of the isomer gamma rays in the $k$th interval of the decay curve is
\begin{equation}
y_k^{is}=N_{is} \frac{q(\lambda_{is}) \Omega}{4 \pi} e^{-\Gamma_{sp} n^3 \big(t_d+(k-1)t_s\big)} \left(1-e^{-t_s \Gamma_{sp} n^3}\right),
\label{eq:5}
\end{equation}
where $q(\lambda_{is})$ is the quantum efficiency of the PMT, $\Omega$ is the solid angle from which the light is collected on the PMT. In our experiment, $\Omega/(4\pi)=0.08$. The delay $t_d$ between the end of excitation and the beginning of detection is 200\,ms. The bin length $t_s$ and number of bins $N_s$ are varied between experimental runs, but the total detection time $t_{\mathrm{det}}=t_sN_s$ is always equal to the excitation time.

\subsection{Expected background count rate}

\subsubsection{Photoluminescence background}

Photoluminescence of Th-doped crystals in response to the excitation light is a major concern, as it may mask the nuclear gamma emission. Previous studies have already characterized the photoluminescence of CaF$_2$ crystals \cite{Mikhailik2006spo,Stellmer2015rap}. In short, the only photoluminescence in CaF$_2$ at timescales longer then 10\,ms after illumination is emitted from self-trapped exciton defects in a well-understood spectrum between 270 and 500\,nm. The residual quantum efficiency of optimized Cs-I photocathodes is a few parts in $10^{-5}$ in this wavelength range. The emission follows a multi-exponential decay for about a minute and then merges into a power-law decay. The amplitude of photoluminescence does not depend on the doping concentration, but on the quality of the crystal.

\subsubsection{Radioluminescence background}

The $^{229}$Th nuclei (half-life 7932\,yrs \cite{Kikunaga2011dot}) decay via a chain of short-lived daughters to $^{209}$Bi. As measured earlier \cite{Stellmer2015rap} and verified in the present study, a 5-MeV alpha decay in CaF$_2$ generates $1.2(2)\times 10^4$ photons between about 225 and 370\,nm. This number does not depend on the quality nor the doping nor the geometry of the crystal. Assuming a light capture efficiency of 8\%, an extra solar blind Cs-I PMT (Hamamatsu R10454) would register 0.03(1) counts for every $^{229}$Th alpha decay (0.15(5) counts for a conventional Cs-I PMT model R8487 and 75(15) counts for a Cs-Te PMT model R8486). There are five alpha decays in the chain of $^{229}$Th daughters. Taking these daughters into account, including a mild scaling for the higher alpha energies, and assuming a crystal with a $^{229}$Th activity of 1\,kBq, we arrive at count rates of 190(60) cps for the extra solar blind Cs-I PMT, five times higher for the normal Cs-I PMT, and $4.8(9) \times 10^5\,$cps for the Cs-Te PMT.

Beta decays in the decay chain of $^{229}$Th generate scintillation light as well, but their contribution is smaller than the uncertainties in the count rates stated above.

The radioluminscence background can be reduced by a factor of about 10 by increasing the crystal temperature from 20$^{\circ}$C to 100$^{\circ}$C \cite{Stellmer2015rap}. The duration of each flash originating from an alpha decay is about 1\,$\mu$s, much shorter than the average time between two alpha decays for a doping of 1\,kBq. An anti-coincidence scheme, building on an additional PMT highly sensitive to the radioluminescence at around 300\,nm, could be implemented to suppress this background entirely.

\subsubsection{Cherenkov radiation background}
\begin{figure}
  \centering
  \includegraphics[width=\columnwidth]{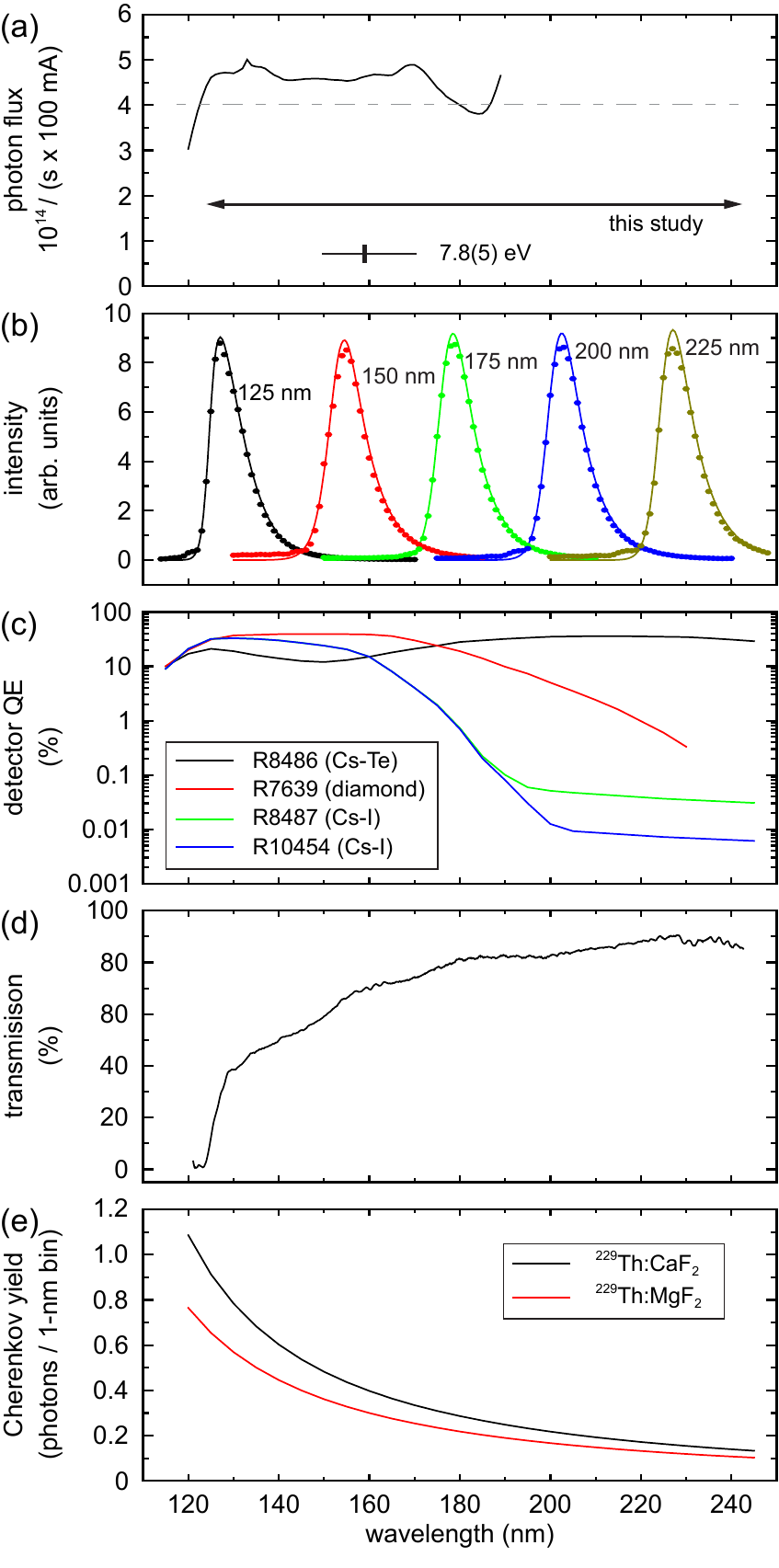}
  \caption{Wavelength-dependent parameters of the experiment. (a) Photon flux on the crystal, measured with a Cs-I photocathode. As a conservative estimate (dashed line), we assume a value of $4\times 10^{14}$ photons per second and 100\,mA beam current across the entire scan range of our study, which covers 124 to 242\,nm. (b) Spectral lineshape of the excitation light measured at five different nominal wavelengths (dots) and fit functions according to Eq.(\ref{eq:excitation_spectrum}) (lines). (c) Quantum efficiencies of the PMT models used in the experiment. (d) Transmission through a 4.1-mm thick CaF$_2$ crystal; adapted from Ref.~\cite{Stellmer2016fso}. (e) Number of Cherenkov photons emitted per disintegration of $^{229}$Th; adapted from Ref.~\cite{Stellmer2016fso}.}
  \label{fig:fig2}
\end{figure}

Cherenkov radiation is caused predominantly by high-energy electrons from beta decays in the $^{229}$Th decay chain. The Cherenkov yield, \textit{i.e.}~the amount of Cherenkov photons emitted per disintegration of a nucleus, can be calculated analytically. We identified the beta decays of $^{213}$Bi and $^{209}$Pb as the major sources of Cherenkov radiation, with much smaller contributions by other beta decays, conversion electrons, and secondary electrons created by Compton scattering from gamma rays. Summing up all contributions along the decay chain of $^{229}$Th, we obtain a value of 0.41 photons emitted in a wavelength band of 1\,nm width centered around 160\,nm \cite{Stellmer2016fso}. The full Cherenkov spectrum is shown in Fig.~\ref{fig:fig2}(e). Multiplying the emission spectrum with the wavelength-dependent quantum efficiency of Cs-I photocathodes, including the transmission of CaF$_2$ crystals, and assuming a collection efficiency of 8\% and a $^{229}$Th activity of 1\,kBq, we arrive at a Cherenkov count rate of 340 cps. For MgF$_2$ as a host matrix, we arrive at a slightly smaller value of 250 cps for the parameters above. Using a diamond photocathode instead, the background counts increase by a factor of two. These are universal numbers for spectroscopy of Th-doped crystals, and may guide the development of protocols for future spectroscopy and clock operation.

Note that we have taken a conservative estimate in that we assumed the Cherenkov radiation to be emitted isotropically. Instead, the radiation is emitted in a narrow cone within a short time of order $\ell/c \approx 10^{-11}\,$s, much shorter than the pulse duration of the PMT. In consequence, the PMT will register either one or zero counts per beta decay, depending on the pointing of the light cone.

Aside from the choice of a crystal with small refractive index, Cherenkov emission is reduced by reducing the lateral size of the crystal to a minimum, such that only a fraction of the electron path is contained in the crystal. The path length of a 1-MeV electron is CaF$_2$ is about 1.7\,mm, much longer than the waist of a tightly focused beam. The crystal diameter would then be matched to the beam.

As Cherenkov emission is the only background radiation in the vicinity of the isomer transition, further measures to reduce this background might need to be implemented. In previous work, we have shown that beta decay in CaF$_2$ is accompanied by substantial scintillation at larger wavelengths, forming a flash of about $10^4$ photons for a 1-MeV beta decay \cite{Stellmer2015rap,Stellmer2016fso}. An auxiliary (bi-)alkali PMT could be used to reliably detect such events, and an anti-coincidence gate could be used to remove the corresponding Cherenkov signal from the signal of the main VUV PMT.

\section{Measurements}
\label{sec:measurements}

The measurements proceed as follows:~the undulator magnet structure is set to generate light at a nominal wavelength $\lambda_0$, and the crystal is illuminated for a time $t_{\rm exc}$. Afterwards, a fast (10\,ms closing time) shutter blocks the undulator radiation from the crystal. Then, a slower beam stop (100\,ms closing time) is moved into the excitation beam to protect the fast shutter from heating by the light. After a delay of 200\,ms following the trigger to the fast shutter, the shutter between the crystal and the PMT is opened (10\,ms opening time), and a trigger is sent to the counting card to commence data acquisition. PMT counts are integrated and saved in bins of 0.1 or 1\,s length, and the number of bins $N_s$ is set such that the detection time $t_{\rm det}=N_s t_s$ equals the excitation time. Afterwards, the nominal wavelength is stepped, the shutters are prepared, and a new excitation/detection cycle starts.

The $1/e$-lifetime of the electron beam in the synchrotron is typically 6 hours, and each measurement run typically lasts 8 hours. Each measurement run seeks to span the sensitivity range of the detector used (typically 80\,nm). The FWHM of the excitation light is about 10\,nm, and we set the typical step size of the wavelength $\Delta \lambda_0$ to 2\,nm. With these parameters, the maximum illumination/detection time is about 10\,min, which limits the range of the isomer's lifetime that can be probed. The measurement campaign described here involves 9 individual measurement runs.

The first measurement run was performed with a Cs-I PMT model R8487, where the wavelength was changed between 124\,nm and 192\,nm in steps of $\Delta \lambda_0=2\,$nm, and the excitation and detection times were set to 120\,s. A large background was observed, originating from the residual sensitivity at long wavelengths.

\begin{figure}
  \centering
  \includegraphics[width=\columnwidth]{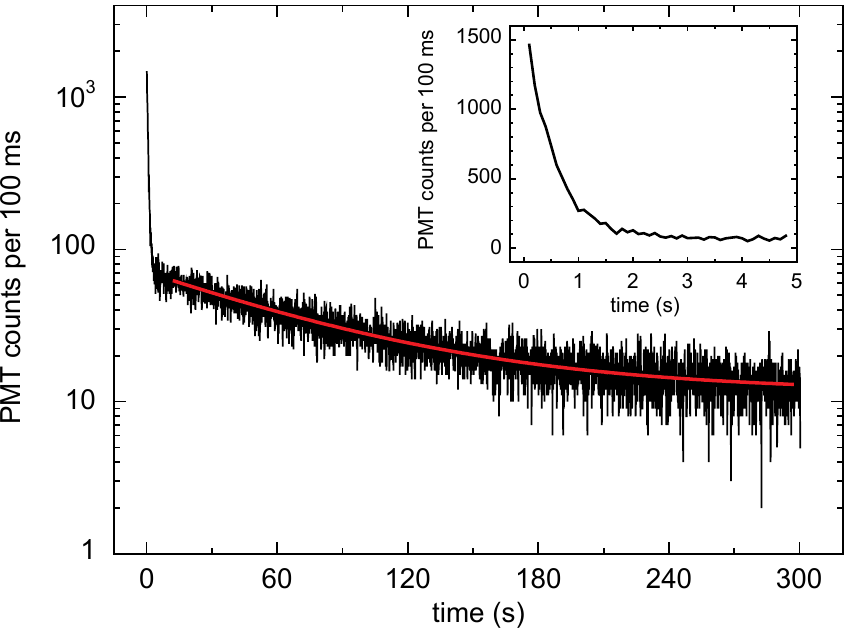}
  \caption{A typical data set showing the light emission of the crystal after excitation. For this specific data set, we used a Cs-I PMT model R10454, an excitation wavelength of $\lambda_0 = 148\,$nm, excitation and detection times of 300\,s, and 100\,ms bin length. An exponential decay fitted to the long-term component is shown as well. Inset: Magnification of the short-term decay on a linear scale.}
  \label{fig:fig3}
\end{figure}

We then performed six measurement runs with the extra solar-blind Cs-I photocathode model R10454, where we used scan ranges of 124\,nm to either 178\,nm or 192\,nm, step sizes of 1 or 2\,nm, bin lengths of 0.1 or 1\,s, and varied the excitation/detection time between 30\,s and 600\,s. The dark count rate of the PMT is 3.2(1)\,cps, and the observed average radioluminescence background is 62(1)\,cps. This value is smaller than the radioluminescence level estimated in the preceding section. We speculate that the residual sensitivity of the PMT at around 300\,nm might be even smaller than specified by the manufacturer, or the crystal might have heated up through absorption of the excitation light, which reduces the radioluminescence. An example curve is shown in Fig.~\ref{fig:fig3}.

We then change to the diamond photocathode and perform two measurement runs in which we change the wavelength in steps of 2\,nm between 124 and 242\,nm. The excitation/detection time is 120\,s, and the bin length is 0.1 and 1\,s, respectively. The dark count rate of the PMT is about 8\,cps, and the radioluminescence background is about 700\,cps. The initial count rates after the end of illumination reach a few $10^5$\,cps, a factor of about 40 larger compared to the Cs-I PMTs. This high sensitivity to the crystal photoluminescence at wavelengths between 270 and 500\,nm puts a limit to the suitability of this model.

In all of these measurement runs, the crystal luminescence is too strong to detect a clear resonance that could be attributed to the isomer decay. Instead, we perform a more elaborate analysis, which will be detailed in the following section. In short, we fit a multi-exponential function to each decay curve, where the decay coefficients are the same for the entire dataset, and only the amplitudes are fit parameters. Afterwards, we check if the fit can be improved by adding an additional contribution from an isomer decay of wavelength $\lambda_{is}$ and decay rate $\Gamma_{sp}$. This comparison then allows us to exclude a certain range of parameters for the isomer.

\section{Data analysis and construction of an exclusion area}
\label{sec:analysis}

\subsection{Selection of data for further analysis}
The expected contribution of the isomer decay to the decay curve is quite weak in comparison to the measured signal caused primarily by crystal photoluminescence. We could not find a highly predictive and accurate quantitative model of the radioluminescence and photoluminescence, therefore, we could detect the manifestation of the isomer transition only if this transition had a quite short lifetime, at the level of one second. Therefore, we focus on measurement runs with high temporal resolution $t_s=0.1\,$s. We take the first 20 points from any decay curve with full resolution, and join the data points between 2 and 30\,s into time bins with 1~s duration each.

\subsection{Fano factors}
The greatest part of the photons reaching the detector originate from photoluminescence and radioluminescence in the crystal. Elementary acts of the relaxation of photoexcitation in the crystal, as well as of the radioactive decay are independent, and their number over a certain time interval obeys Poissonian distribution. At the same time, each of such an act may be accompanied by non-simultaneous emission of several photons which, being detected, cause several counts in the photomultiplier. Therefore, the number of PMT counts over a certain time interval is no longer Poissonian, but {\em super-Poissonian}. The dispersion of the number of counts per time interval is proportional to the mean, but the coefficient of proportionality, the {\em Fano factor}, is larger than 1.

We have chosen 3 runs with high temporal resolution for detailed analysis, one of them was performed with a diamond photocathode (model R7639), and another two with an extra solar-blind Cs-I photocathode (model R10454). The Fano factor will depend on the type of photocathode, because elementary events in the crystals are accompained primarily by longer-wavelength photons, and the diamond photocathode is more sensitive to longer wavelength photons than the Cs-I model; see Fig.~\ref{fig:fig2}(c). A preliminarly estimation of these Fano factors with the help of segmentized-parabolic fits gives $F_{\mathrm{dia}}\approx 1.6 - 1.7$ and $F_{\mathrm{CsI}}\approx 1 - 1.2$.

\subsection{Parametrization of decay curves}
As a basic model of the decay curves, we consider multi-exponential models with decay constants $\gamma_\alpha$ common for all the decay curves, but amplitude coefficients $C_{i,\alpha}$ specific for each curve. Negative values of $C_{i,\alpha}$ are allowed. Also, we suppose the presence of a time-independent term reflecting radioluminescence and detector dark counts. This term can be included into the multi-exponential model with a decay constant $\gamma_0=0$. Therefore, the number $n_{i,k}$ of counts in the $k$th interval of the $i$th decay curve is a (super-Poissonian) random value with mean 
\begin{equation}
y_{i,k}^{0}=\sum_{\alpha}C_{i,\alpha} \exp[-t_k \gamma_\alpha] \frac{1-\exp[-\gamma_\alpha \, \Delta t_k]}{\gamma_\alpha}.
\label{eq:7}
\end{equation}
The superscript in $y_{i,k}^{0}$ indicates that this model includes no isomer decay. Here, $t_k$ is the beginning of $k$th time interval, and $\Delta t_k$ is its duration. For $\gamma_{\alpha}=0$, the fraction in Eq.\,(\ref{eq:7}) should be replaced by its limit value $\Delta t_{k}$.

We will now introduce a possible isomer decay by adding another term to the model,
\begin{equation}
\begin{split}
\lefteqn{y_{i,k}^1(\lambda_{is},\Gamma_{sp},I_i)=  y_{i,k}^{is}(\lambda_{is},\Gamma_{sp},I_i)} \\ & +\sum_{\alpha}C_{i,\alpha} \exp[- t_k \gamma_\alpha]\frac{1-\exp[-\gamma_\alpha \, \Delta t_k]}{\gamma_\alpha},
\end{split}
\label{eq:8}
\end{equation}
where the superscript $1$ in $y_{i,k}^1$ indicates the presence of the isomer decay, $y_{i,k}^{is}(\lambda_{is},\Gamma_{sp},I_i)$ is given by Eq.\,({\ref{eq:5}}), and $N_{Th,is}$ is taken to be equal to $N_{eq}$ calculated according to Eq.\,({\ref{eq:4}}), because we consider here only short isomer lifetimes, much shorter than the time of excitation. This model has two additional fit parameters, namely the wavelength $\lambda_{is}$ and the (bare) decay rate $\Gamma_{sp}$ of the isomer transition; see Eq.\,({\ref{eq:5}}).

The idea of this analysis is to find the best fit according to model (\ref{eq:8}), and to determine an {\em acception region} in the plane $(\lambda_{is},\Gamma_{sp})$, which includes the real values of the isomer wavelength and decay rate. To characterize different fits corresponding to different models and values of $\lambda_{is}$ and $\Gamma_{sp}$, we introduce the value $X(\lambda_{is},\Gamma_{sp})$, the sum of normalized squared differences between the number of PMT counts $n_{i,k}$ and its fitted expectation $y_{i,k}$ corresponding to model (\ref{eq:8}),
\begin{align}
X(\lambda_{is},\Gamma_{sp}) =&\sum_{i} \frac{1}{F_{i}} \sum_{k} \frac{(y^{1}_{i,k}-n_{i,k})^2}{y^*_{i,k}}.
\label{eq:9} 
\end{align}
Here, the index $i$ runs over decay curves, $k$ runs over time intervals of any decay curve, $\alpha$ runs over decay constants, and $N_{i}$ is the number of time intervals in the $i$th curve (in our case, $N_i=20$ for all $i$). The normalization function $y^{*}_{i,k}$ should approximate the true (unknown) expectation number of counts $n_{i,k}$; we use the best fit $y^{0}_{i,k}$ according to our basic model (\ref{eq:7}) as $y^{*}_{i,k}$. 

If the model (\ref{eq:8}) is correct, and the numbers of counts $n_i$ are super-Poissonian random values with dispersion $F_{i} y^{*}_{i,k}$, then the value $X(\lambda_{is}^*,\Gamma_{sp}^*)$, where $\lambda_{is}^*$ and $\Gamma_{sp}^*$ are the (unknown) true values of the isomer transition wavelength and the isomer decay rate in the bare nucleus, respectively, should approximately be a $\chi^2_S$ random value with $S$ degrees of freedom. Here, $S$ is the total number of time intervals minus the number of free parameters, \textit{i.e.}, minus the total number of coefficients $C_{i,a}$ and fitted decay constants $\gamma_{\alpha}$. Note that we use the same decay constants but independent coefficients $C_{i,a}$ for all curves, so the number of coefficients $C_{i,a}$ is much larger than the number of fitted constants $\gamma_{\alpha}$. If we treat the parameters $\lambda_{is}$ and $\Gamma_{sp}$ as free parameters as well, and fit their best values $\lambda_{is}^b$ and $\Gamma_{sp}^b$, then the difference
\begin{align}
\Delta X=X(\lambda_{is}^*,\Gamma_{sp}^*)-X(\lambda_{is}^b,\Gamma_{sp}^b)
\label{eq:10} 
\end{align}
is approximately a $\chi^2$-random value with two degrees of freedom \cite{Rolke2001cia,Casella2001sti}. It allows us to determine the confidence region for parameters $(\lambda_{is},\Gamma_{sp})$.

\subsection{Fano factors and choice of time constants}

To compare different fits with the help of Eqs.\,(\ref{eq:9}) and (\ref{eq:10}), we first estimate the Fano factors $F_{\mathrm{dia}}$ and $F_{\mathrm{CsI}}$, the number of fitted decay constants $\{\gamma_{\alpha}\}$, and fix some of these constants. For a preliminarly estimation of the time constants we used the basic model (\ref{eq:7}).

At first, we consider the long-term part of the decay curves for $10~{\rm s}<t<30~{\rm s}$, which shows a long-lived component; see Fig.~\ref{fig:fig3}. The self-consistent estimation is performed in the following manner:~firstly, we take $F_{i}=1$ and fit the value of $\gamma_L$ for the data obtained with the diamond and Cs-I photocathodes independently using model (\ref{eq:7}) with one decaying component and one time-independent term. For both PMTs, the best values of $\tau_L=1/\gamma_L$ are slightly above 1\,min. We then calculate the values of $X$ according to Eq.\, (\ref{eq:10}), and estimate the respective Fano factors as $X/S$, where $S$ is the number of points minus the number of free parameters. Further, using these estimations of both  Fano factors, we vary the value $\gamma_{L}$ of decay constants, trying to minimize the value of $X$ calculated for all  decay curves. Using this minimized value of $X$, we correct our estimation of $F_{\mathrm{dia}}$ and $F_{\mathrm{CsI}}$ in such a way that the value of $X$ calculated for each PMT is equal to $S$. This procedure is performed several times and gives a self-consistent value of $\tau_{L}=1/\gamma_{L}=68.75$~s, and Fano factors $F_{\mathrm{dia}}=1.627$ and $F_{\mathrm{CsI}}=1.092$. The value of $\tau_L$ is not varied further.

In the next step, we consider all decay curves and perform a similar self-consistent procedure to fit the data with the help of the basic model (\ref{eq:7}) with additional short-term decay components. When allowing for only one short-term decay component, we recover the value of $\tau = 0.45\,{\rm s}$ reported in our earlier work \cite{Stellmer2015rap}. The optimal fits contains three short-term decay components $\tau_{\alpha}=1/\gamma_\alpha$:
\begin{equation}
\tau_1=0.343~{\rm s}; \quad \tau_2=0.540~{\rm s}; \quad \tau_3=1.467~{\rm s}.
\label{eq:14}
\end{equation}
The corrected estimation of the Fano factors are $F_{\mathrm{dia}}=1.6135$ and $F_{\mathrm{CsI}}=1.0725$. These values are consistent with the ones obtained above for $10~{\rm s}<t<30~{\rm s}$. The best fit with two short-term components instead of three gives a too high value of $X$, inconsistent with the abovementioned estimations of Fano factors and corresponding numbers of degrees of freedom. We decide to use three variable short-term decay constants $\gamma_\alpha$ ($\alpha=1,2,3$), one fixed long-term decay constant $\gamma_{L}=1/68.75~{\rm s^{-1}}$, and one time-independent term. We used the basic model (\ref{eq:7}) for this procedure, because the model (\ref{eq:8}) with the isomer signal depends on the unknown values of $\lambda_{is}$ and $\Gamma_{sp}$. Moreover, single decay curves calculated according to models (\ref{eq:7}) and (\ref{eq:8}) differ only slightly. For the same reason, we use the best fit $y^{0}_{i,k}$ as $y^{*}_{i,k}$.

\begin{figure}
  \centering
  \includegraphics[width=\columnwidth]{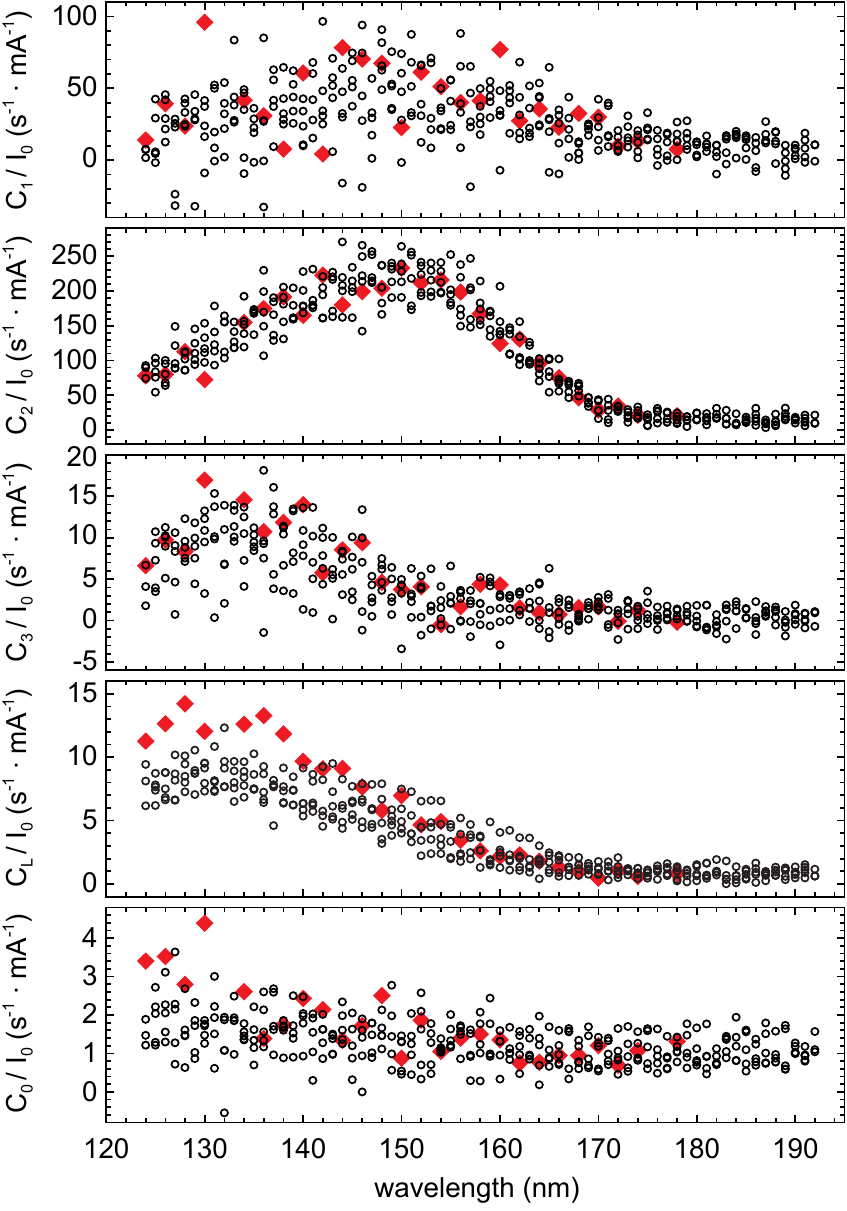}
  \caption{Coefficients $C_{i,\alpha}$ for various decay curves corresponding to the best fits with model (\ref{eq:7}) with 3 non-zero short-term decay constants $\gamma_{\alpha}$ given in (\ref{eq:14}), one long-term component $\gamma_L$, and one time-independent term, as a function of the nominal wavelength $\lambda_0$ of the excitation light. The example curves shown here are obtained with a Cs-I photocathode, where red diamonds (black circles) show a dataset with time bins of 1\,s (100\,ms).}
  \label{fig:fig4}
\end{figure}

The coefficients $C_{i,1}$, $C_{i,2}$, $C_{i,3}$ in Eq.\,(\ref{eq:7}) corresponding to short-term decay rates (\ref{eq:14}), as well as the coefficients $C_{i,L}$ of the long-term component and $C_{i,0}$ of the time-independent background are presented in Fig.~\ref{fig:fig4}

\subsection{Fit and confidence region for parameters of the isomer transition}

We perform a series of fits with coefficients $C_{\alpha}$ and short-term decay constants $\gamma_1, \gamma_2, \gamma_3$ according to model (\ref{eq:8}) with an assumed isomer signal for different values of $\lambda_{is}$ (between 115 and 240 nm) and $\Gamma_{sp}$ (between 0.4 and $10\,\mathrm{s}^{-1}$) at a fixed value of the long-term decay constant $\gamma_L$, seeking to minimize the value of $X(\lambda_{is},\Gamma_{sp})$ calculated according to (\ref{eq:9}). Considering now $\lambda_{is}$ and $\Gamma_{sp}$ as fit parameters, we find a minimal value $X(\lambda_{is}^b,\Gamma_{sp}^b)=X_0-1.1243$ at ($\lambda_{is}^b=160$~nm, $\Gamma_{sp}^b=0.6~{\rm s^{-1}}$), where $X_0=22661$ is calculated for the basic model (\ref{eq:7}). This small change suggests that this pair of parameters cannot be assigned to the isomer. In other words, the confidence region is an open set on the $(\lambda_{is},\Gamma_{sp})$-plane.

\begin{figure}
  \centering
  \includegraphics[width=\columnwidth]{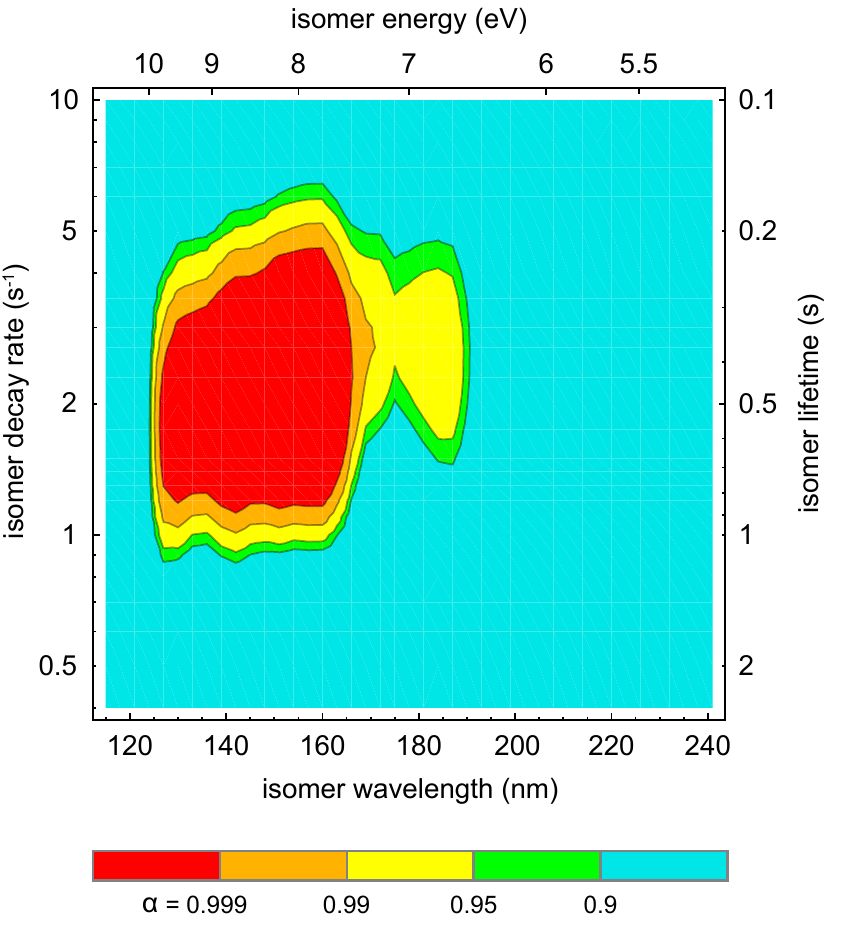}
  \caption{A contour plot of the parameter $\alpha$ in the plane of isomer wavelength $\lambda_{is}$ and bare decay rate $\Gamma_{sp}$. The parameter $\alpha$ describes the level at which a pair of parameters $(\lambda_{is},\Gamma_{sp})$ is \textit{excluded} by our measurements, see the text for details. To give an example, the region shaded in red can be excluded at the 99.9\% level.}
  \label{fig:fig5}
\end{figure}

Any pair $(\lambda_{is},\Gamma_{sp})$ may be characterized by the value
\begin{equation}
\alpha(\lambda_{is},\Gamma_{sp})=P\big[\chi^2_2 < X(\lambda_{is},\Gamma_{sp})-X(\lambda_{is}^b,\Gamma_{sp}^b)\big],
\label{eq:15}
\end{equation}
where $P\big[\chi^2_2 < \Delta X\big]$ is the probability that the $\chi^2$-random value is less than $\Delta X$. If $\alpha(\lambda_{is},\Gamma_{sp})<c$, we assign the pair $(\lambda_{is},\Gamma_{sp})$ to the confidence region corresponding to significance level $c$. For conventional values of the significance level ($c=0.9$ to 0.999), the confidence region is open in the $(\lambda_{is},\Gamma_{sp})$-plane. This map of $\alpha(\lambda_{is},\Gamma_{sp})$ is presented in Fig.~{\ref{fig:fig5}} and visualizes the region of parameters that is \emph{excluded} by our experiment. This finding is the main result of our study.

\section{Conclusion}
\label{sec:conclusion}

In the work presented here, we have, for the first time, covered a spectral region of $\pm5\,\sigma$ around the expected isomer energy of 7.8(5)\,eV in optical excitation of doped crystals using synchrotron radiation. Three detectors with different spectral sensitivities were used, and the excitation time was varied between 30 and 600 seconds. We found an unexpectedly strong photoluminescence background emitted by the crystal in response to the excitation light. As a consequence, the experiment could detect gamma emission of $^{229\mathrm{m}}$Th only if the isomer lifetime was on the order of 1\,s. When assuming 100\% radiative de-excitation, the data allows us to exclude an isomer with energy and lifetime between 7.5 and 10\,eV and 0.2 and 1.1\,s, respectively. This result extends the parameter region excluded by Jeet \textit{et~al.} considerably \cite{Jeet2015roa,Tkalya2015rla} and is in agreement with recent studies by von der Wense \textit{et~al.} \cite{Wense2015ddo}.

The crystal's photoluminescence appears only at wavelengths above 270\,nm \cite{Stellmer2015rap}, yet residual sensitivity even of solar-blind VUV photocathodes in this wavelength range leads to a massive count rate which masks the isomer signal. In future work, we will grow crystals of higher purity to reduce the photoluminescence yield substantially. This could be achieved with MgF$_2$ crystals, which are less hygroscopic compared to CaF$_2$. Inserting a grating into the optical pathway towards the detector would rigorously separate crystal scintillation from the isomer emission, at the same time reducing the light throughput by two orders of magnitude. Alternatively, a light source with smaller linewidth needs to be employed. 

Cherenkov radiation, caused by unavoidable beta activity in the crystal, leads to a background at the same wavelength as the sought-after nuclear gamma emission. The amplitude of this background has been calculated. It can be reduced by choice of a crystal with smaller refractive index (\textit{e.g.}~MgF$_2$), and by a reduction of the crystal's lateral extent. Based on our numbers, one can calculate the minimum spectral power of a light source required to observe the isomer emission for a given isomer lifetime. Ultimatively, we suggest the implementation of an anti-coincidence scheme that builds on the detection of scintillation light at longer wavelengths.

Crystal scintillation in response to alpha decay appears at a wavelength around 300\,nm, where VUV photocathodes show a residual senstivity of order $10^{-5}$. The resulting count rates constitute a considerable background, which was quantified in this work and can be suppressed by an anti-coincidence scheme as well.

Lastly, we have developed and successfully tested elaborate data processing software to analyze the spectroscopic data. In a future measurement campaign, we will employ Th-doped MgF$_2$ crystals and probe substantially longer isomer lifetimes. The optical excitation of $^{229\mathrm{m}}$Th remains an exquisite challenge.

\section*{Acknowledgements}
We thank E.~Peik, G.~Ulm, A.~Gottwald, Y.~Shigekawa, L.~von der Wense, B.~Seiferle, and P.G.~Thirolf for fruitful discussions. We acknowledge experimental support by J.~Sterba, V.~Rosecker, A.~Sas, and M.~Fritthum. This project has received funding from the European Union’s Horizon 2020 research and innovation programme under grant agreement No.~664732, as well as from the Austrian FWF SFB project ViCoM.

\bibliographystyle{apsrev}
\bibliography{bib}

\end{document}